\documentclass[a4]{article}

\usepackage{amssymb,amsmath,amsthm}
\usepackage{newtxtext} % times
\usepackage[scaled=.95]{cabin} % sans serif
\usepackage[varqu,varl]{inconsolata} % typewriter
\usepackage{amsmath}
\usepackage[varg]{newtxmath}
\usepackage{ifxetex,ifluatex}
\usepackage{fixltx2e} % provides \textsubscript
\usepackage{algorithm2e}
\ifnum 0\ifxetex 1\fi\ifluatex 1\fi=0 % if pdftex
  \usepackage[T1]{fontenc}
  \usepackage[utf8]{inputenc}
\else % if luatex or xelatex
  \ifxetex
    \usepackage{mathspec}
  \else
    \usepackage{fontspec}
  \fi
  \defaultfontfeatures{Ligatures=TeX,Scale=MatchLowercase}
\fi
\IfFileExists{upquote.sty}{\usepackage{upquote}}{}
\usepackage{microtype}
\UseMicrotypeSet[protrusion]{basicmath} % disable protrusion for tt fonts
\usepackage[margin=1in]{geometry}
\usepackage{hyperref}
\hypersetup{unicode=true,
            pdftitle={The reference model approach in variable selection},
            pdfauthor={Federico Pavone, Juho Piironen, Paul-Christian B\"{u}rkner, Aki Vehtari},
            pdfkeywords={keywords},
            pdfborder={0 0 0},
            breaklinks=true,
            colorlinks=true,
            linkcolor=black,
            citecolor=RoyalBlue,
            filecolor=black,
            urlcolor=RoyalBlue,
          }
\ifnum 0\ifxetex 1\fi\ifluatex 1\fi=0 % if pdftex
  \usepackage[shorthands=off,main=american]{babel}
\else
  \usepackage{polyglossia}
  \setmainlanguage[variant=american]{english}
\fi
\usepackage{natbib}
\usepackage{placeins}
\usepackage{graphicx,grffile,subcaption}
\makeatletter
\def\maxwidth{\ifdim\Gin@nat@width>\linewidth\linewidth\else\Gin@nat@width\fi}
\def\maxheight{\ifdim\Gin@nat@height>\textheight\textheight\else\Gin@nat@height\fi}
\makeatother
\setkeys{Gin}{width=\maxwidth,height=\maxheight,keepaspectratio}
\IfFileExists{parskip.sty}{%
\usepackage{parskip}
}{% else
\setlength{\parindent}{0pt}
\setlength{\parskip}{6pt plus 2pt minus 1pt}
}
\ifx\paragraph\undefined\else
\let\oldparagraph\paragraph
\renewcommand{\paragraph}[1]{\oldparagraph{#1}\mbox{}}
\fi
\ifx\subparagraph\undefined\else
\let\oldsubparagraph\subparagraph
\renewcommand{\subparagraph}[1]{\oldsubparagraph{#1}\mbox{}}
\fi

\let\rmarkdownfootnote\footnote%
\def\footnote{\protect\rmarkdownfootnote}

\usepackage{titling}

\usepackage{booktabs}
\usepackage{longtable}
\usepackage{array}
\usepackage{multirow}
\usepackage[table,dvipsnames]{xcolor}
\usepackage{wrapfig}
\usepackage{float}
\usepackage{colortbl}
\usepackage{pdflscape}
\usepackage{tabu}
\usepackage{threeparttable}
\usepackage{threeparttablex}
\usepackage[normalem]{ulem}
\usepackage{makecell}

\usepackage{soul}
\usepackage{mathtools}
\usepackage{textcomp}
\usepackage{graphicx,pdflscape}
\usepackage{geometry}
\usepackage{amsmath}
\usepackage{float}
\usepackage{supertabular}
\usepackage{booktabs,fixltx2e}
\usepackage[font={small,it}]{caption}
\usepackage{tcolorbox}
\usepackage{paralist}
\usepackage{multicol}
\setcitestyle{round}

\theoremstyle{definition}

\DeclareMathOperator{\Cor}{Cor}
\DeclareMathOperator{\N}{N}

\title{Using reference models in variable selection 
	\vspace{.1in}}
    \pretitle{\vspace{\droptitle}\centering\huge}
  \posttitle{\par}
  \author{Federico Pavone, Juho Piironen, Paul-Christian B\"{u}rkner and Aki Vehtari}
    
    \author{
      Federico Pavone\footnote{Department of Computer Science, Aalto University, Finland.}\, \footnote{Department of Decision Sciences, Bocconi University, Italy.}%$\,$\footnote{Correspondence concerning this article should be addressed to Federico Pavone, Department of Decision Sciences, Bocconi University, Via R\"{o}ntgen 1, 20136 Milano, Italy. E-mail: federicopavone01@gmail.com.}
      \,,\,
  Juho Piironen$^*$,\,
  Paul-Christian B\"{u}rkner$^*$
  and Aki Vehtari$^*$
  }

    \preauthor{\centering\large\emph}
  \postauthor{\par}
    \date{\today}
    \date{\today}
\begin{document}
\maketitle
\begin{abstract}
  Variable selection, or more generally, model reduction is an important aspect of the statistical workflow aiming to provide insights from data. In this paper, we discuss and demonstrate the benefits of using a reference model in variable selection. A reference model acts as a noise-filter on the target variable by modeling its data generating mechanism. As a result, using the reference model predictions in the model selection procedure reduces the variability and improves stability leading to improved model selection performance. Assuming that a Bayesian reference model describes the true distribution of future data well, the theoretically preferred usage of the reference model is to project its predictive distribution to a reduced model leading to projection predictive variable selection approach. Alternatively, reference models may also be used in an ad-hoc manner in combination with common variable selection methods.  In several numerical experiments, we investigate the performance of the projective prediction approach as well as alternative variable selection methods with and without reference models. Our results indicate that the use of reference models generally translates into better and more stable variable selection. Additionally, we demonstrate that the projection predictive approach shows superior performance as compared to alternative variable selection methods independently of whether or not they use reference models.
\end{abstract}

\hypertarget{introduction}{%
\section{Introduction}\label{introduction}}

In statistical applications, one of the main steps in the modelling
workflow is variable selection, which is a special case of
model reduction. Variable selection (also known as feature or 
covariate selection) may have multiple goals.  First,
if the variables themselves are of interest, we can use variable
selection to infer which variables contain predictive information
about the target. Second, as simpler models come with the
advantages of reduced measurement costs and improved interpretability,
we may be interested in finding the minimal subset of variables which
still provides good predictive performance (or good balance between
simplicity and predictive performance).  When the predictive
capability is guiding the selection, the true data generation
mechanism of future data can be approximated either by using the
observed data directly or alternatively by using predictions from a
\emph{reference model} \citep{vehtari2012survey}.

In data-based approaches, such as Lasso selection
\citep{tibshirani1996regression} or stepwise backward/forward
regression \citep{venables2013modern,harrell2015regression}, the
observed empirical data distribution is utilised as a proxy of future
data usually in combination with cross-validation or information
criteria to provide estimates for out-of-sample predictive
performance.  In contrast, reference model based methods approximate
the future data generation mechanism using the predictive distribution
of a reference model, which can be, for example, a full-encompassing
model including all variables.  The main assumption under any
reference model approach is that we operate in an
$\mathcal{M}$-\textit{complete} framework
\citep{book:bernardo_smith,vehtari2012survey}, that is, we have
constructed a model which reflects our beliefs about
the future data in the best possible way and which has passed model checking and criticism
\citep[see, e.g.][]{gelman2013bayesian,gabry2019visualization}.  The reference model approach
has been used in Bayesian statistics at least since the seminal
work of \citet{paper:reference_lindley}. For more historical
references, see \citet{vehtari2012survey} and
\citet{paper:model_selection}, and for most recent methodological
developments see \citet{paper:projpred}.

Reference models have been also used in non-Bayesian contexts, in which
\cite{harrell2015regression} describes them as full models that can be
thought of as a ``gold standard'' (for a given application).  For
example, \cite{faraggi2001understanding} deal with the necessity of
identifying interpretable risk groups in the context of survival data
using neural networks, which typically perform very well in terms of
prediction, but whose variables are difficult to be understood in
terms of relevance.  \cite{paul2008preconditioning}, using the term
preconditioning, explore approximating models fitting Lasso or
stepwise regression against consistent estimates $\hat{y}$ of a
reference model instead of the observed responses $y$.
Whatever the terminology or applied statistical framework, reference
models offer a powerful approach to improving variable selection as we
will demonstrate in the present paper.

This paper is structured as follows. In Section
\ref{ref-intro}, we review the concept of the reference
model, its benefits with examples and how it can be used as a filter
on data in a simple way. In Section \ref{comparison-minimal-subset} and 
Section \ref{comparison-complete-subset}, we show the
benefits of a reference model approaches for minimal and complete 
variable selection, respectively, before we end with a conclusion in Section \ref{conclusion}.

\section{Reference models in variable selection}
\label{ref-intro}

In this section, we will provide an initial motivation and intuition 
for the use of reference models to improve variable selection methods.
We will start with a case study that is repeatedly used throughout
the paper to illustrate the benefits of reference models before we dive deeper
into the theoretical reasons why reference models help in variable selection.

\subsection{Body fat example: Part 1}
\label{bodyfat-1}

To motivate the further discussion and experiments, we start by a simple
variable selection example using body fat data by \citet{johnson1996fitting}.
We compare the projective prediction approach (\emph{projpred},
\citet{paper:projpred}) which uses a reference model, and classic
stepwise backward regression (\emph{steplm}).
The experiments are implemented in \texttt{R} \citep{Rcore2018}.

The target variable of interest is the amount of body fat, which is
obtained by a complex and expensive procedure consisting in immersing a 
person in a water tank and carrying out different measurements and computations. 
Additionally, we have information about 13
variables which are anthropometric measurements (e.g., height, weight
and circumference of different body parts). The
variables are highly correlated which causes additional
challenge in the variable selection. In total, we have 251
observations. The goal is to find the model which is able to predict
the amount of body fat well while requiring the least amount of
measurements for a new person.

\cite{paper:bodyfat} report results using steplm with a significance
level of 0.157 with AIC selection \citep{akaike1974new}, fixing
abdomen and height to be always included in the model. For better
comparison we do not fix any of the variables.  The steplm approach is
carried out combining the \texttt{step} and \texttt{lm} functions in \texttt{R}.

For the selection via projpred, the Bayesian reference model includes
all the variables using a regularised horseshoe prior~\citep{paper:rhs} on the
variable coefficients. Submodels are explored using
forward search (the results are not sensitive to whether forward or
backward search is used), and the predictive utility is the expected
log-predictive density (elpd) estimated using approximate leave-one-out cross-validation via Pareto-smoothed importance-sampling
\citep[PSIS-LOO-CV; ][]{paper:psis_loo}.  We select the smallest  submodel with an elpd score similar to the reference model
when taking into account the uncertainty in estimating the predictive
model performance. See Appendix A for brief review of projection
predictive approach, and papers by \citet{paper:model_selection} and
\citet{paper:projpred} for more details. The complete projpred approach is implemented in the \texttt{projpred} \texttt{R} package \citep{Rprojpred}.

\begin{figure}[tp]
  \centering
  \includegraphics[width=0.98\textwidth]{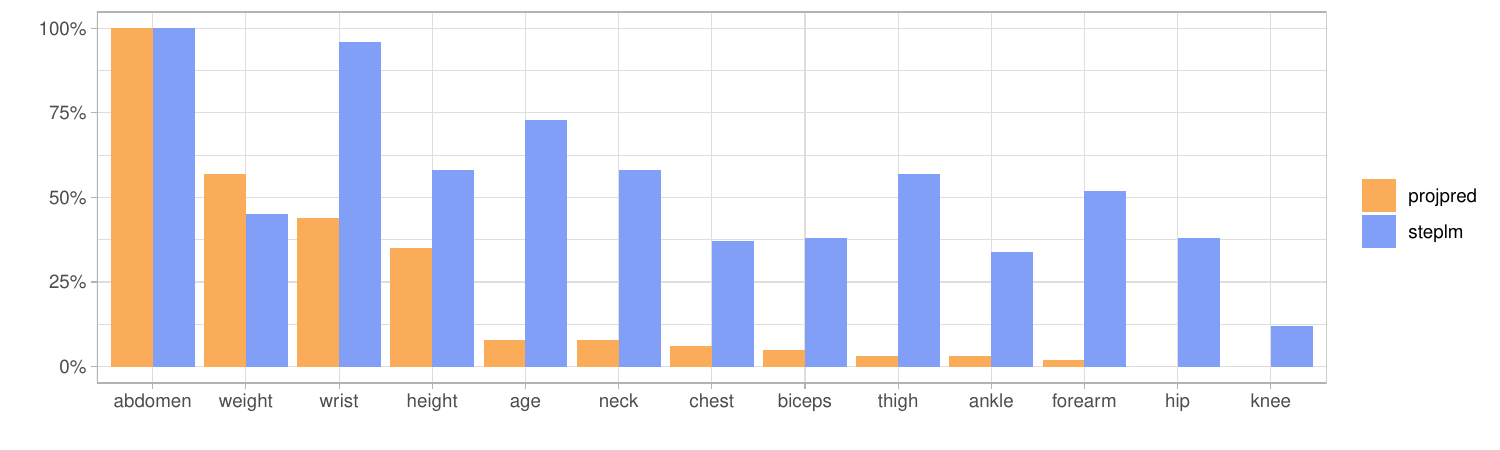}
  \vspace{-0.7\baselineskip}
  \caption{Body fat example: Bootstrap inclusion frequencies calculated from 100 bootstrap samples. The projpred approach has less variability on which variables are selected.}
  \label{fig:inclusion_frequencies}
\end{figure}

The inclusion frequencies of each variable in the final model given 100
bootstrap samples are shown in Figure
\ref{fig:inclusion_frequencies}. In case of projpred there are two
variables, `abdomen' and `weight', which have inclusion frequencies
above 50\% (`abdomen' is the only one included always), the third most
frequently included is `wrist' at 44\%, and the fourth one is `height'
at 35\%.  The steplm approach has seven variables with inclusion
frequencies above 50\%. Such a higher variability and lower stability
of steplm can be observed also in the bootstrap model selection
frequencies reported in Table \ref{tab:model_frequencies}. For
example, the first five selected models have a cumulative frequency of
76\% with projpred, but only of 14\% with steplm. In addition, the
sizes of the selected models with projpred are much smaller than the ones 
selected with steplm.

\begin{table}[tp]
\footnotesize
\centering
\begin{tabular}{l|lr|lr}
M & projpred & Freq \% & steplm & Freq \%  \\ 
  \hline
1 & abdom., weight & 39 & abdom., age, forearm, height, hip, neck, thigh, wrist & 4 \\
2 & abdom., wrist & 10 & abdom., age, chest, forearm, height, neck, thigh, wrist & 4 \\
3 & abdom., height & 10 & abdom., forearm, height, neck, wrist & 2 \\
4 & abdom., height, wrist & 9 & abdom., forearm, neck, weight, wrist & 2 \\
5 & abdom., weight, wrist & 8 & abdom., age, height, hip, thigh, wrist & 2 \\
6 & abdom., chest, height, wrist & 2 & abdom., age, height, hip, neck, thigh, wrist & 2 \\
7 & abdom., biceps, weight, wrist & 2 & abdom., age, ankle, forearm, height, hip, neck, thigh, wrist & 2 \\
8 & abdom., height, weight, wrist & 2 & abdom., age, biceps, chest, height, neck, wrist & 2 \\
9 & abdom., age, wrist & 2 & abdom., age, biceps, chest, forearm, height, neck, thigh, wrist & 2 \\
10 & abdom., age, height, neck, thigh, wrist & 2 & abdom., age, ankle, biceps, weight, wrist & 2 \\
\end{tabular}
\caption{Body fat example: Bootstrap model selection frequencies from 100 bootstrap samples.  The projpred approach has less variability on which variable combinations are selected.}
\label{tab:model_frequencies}
\end{table}

The first two rows of Table \ref{tab:model_performances} show the
predictive performances, in terms of cross-validated root mean square
error (RMSE), of the full model and the selected models using projpred
or steplm.  There is no significant difference in predictive performance
of the selected models by different approaches even if there is
clear difference in the number of selected variables. This can be explained by high
correlation between the variables and different combinations can
provide similar predictive accuracy.

We repeat the experiment with a modified data set by adding 84
unrelated noisy variables, resulting in 100 variables in total. 
The last two rows of Table
\ref{tab:model_performances} show the cross-validated RMSE, the size
of the selected model and the number of selected noisy variables using
projpred or steplm. The results show that projpred has similar
predictive performance and the same number of selected variables as
with the original data, whereas the stepwise regression has worse
predictive performance and the number of selected variables is much
higher and include a large number of irrelevant variables.

\begin{table}[tp]
\footnotesize
\centering
\begin{tabular}{ll|rrrr}
Data  & Method & RMSE Full & RMSE Sel & \# Sel (10-CV avg, sd) & \# Sel noisy (10-CV avg, sd) \\ 
  \hline
body fat & projpred & 4.4 & 4.5 &  2\, (2.3, 0.5) &   \\
& steplm & 4.4 & 4.5 & 7\, (6.0, 0.9) &  \\
\hline
+ noisy variables & projpred & 4.5 & 4.5 &  2\, (2.0, 0.0)&  0\,\; (0,\, 0)  \\
& steplm & 5.7 & 5.1 & 23\, (26,\; 4.5) &  15\, (19, 4) \\
\end{tabular}
\caption{Body fat example: Predictive performances with original data (first two
  rows) and with extra noisy variables (last two rows) estimated with
  10-fold cross-validation. Abbreviations: RMSE = root mean squared error;
  Full = full model; Sel = selected submodel; 
  \# Sel = total number of selected variables;
  \# Sel noisy = number of selected noisy variables. 
  10-CV avg = average over the 10 folds in cross-validation. 10-CV sd = standard deviation
  over the 10 folds in cross-validation.}
\label{tab:model_performances}
\end{table}

Both projpred and steplm compare a large number of models using either
forward or backward search, which can lead to selection induced
overfitting, but even with 100 variables, projpred is able to select a
submodel with similar performance as the full model. In this example, the two
 compared methods also differ in other aspects than the usage of a reference 
 model, such as that projpred
uses Bayesian inference and steplm uses maximum likelihood estimation. 
However, as we show in the present paper, one of the primary difference 
with respect to quality of the variable selection is indeed whether or 
not a reference model is applied.

\subsection{Benefits and costs of using a reference model}

A properly designed reference model is able to filter parts of the
noise present in the data, and hence to provide an improved and more
stable selection process. This holds even if the reference model does
not perfectly resemble the true data generating process. Moreover, our
analyses indicate that the substantial reduction of variance
attributable to noise is usually more important than small potential
bias due to model misspecification.  We argue that, regardless of how
the reference model is set up and used in the inference procedure, it
can be always seen as acting as a filter on the observed
data. Furthermore, regardless of what specific model selection method
is used, a reference model can be used instead of raw data during the
selection process to improve the stability and selection
performance. Our results indicate that the core reason why the
reference model based methods perform well is the reference model
itself, rather than the specific way of using it.  In general, the
less data we have and the more complex the estimated models are, the
higher is the benefit of using a reference models as the basis for
variable selection.

If one of the models to be compared is the full model, which can be
used as a reference model, there is no additional cost of using a
reference model as it was estimated as part of the analysis anyway.
Sometimes including all the available variables in an
elaborate model can be computationally demanding.  In such a
case, even simpler screening or dimensionality reduction techniques,
as for example the supervised principal components
\citep{bair2006prediction,piironen2018} can produce useful reference
models.

\hypertarget{reference-model-approach}{
\subsection{Why the reference model helps}\label{reference-model-approach}}
 
A good predictive model is able to filter part of the noise present in
the data. The noise is the main source of the instability in the selection
and tends to
obscure the relevance of the variables in relation to the target
variable of interest. We demonstrate it with the following simple
explanatory example taken from \cite{paper:projpred}. The data
generation mechanism is
\begin{alignat}{2} \label{eq:simulated_data}
     &f\sim \N(0,1) && \nonumber \\ 
     Y|&f\sim \N(f,1) && \\
     X_{j}|&f \overset{iid}{\sim} \N(\sqrt{\rho}f,1-\rho) \quad &&j=1,\ldots,k \nonumber \\
     X_{j}|&f \overset{iid}{\sim} \N(0,1) &&j=k+1,\ldots,p \nonumber,
\end{alignat}
where $f$ is the latent variable of interest of which $Y$ is a noisy
observation. The first $k$ variables are strongly related to the
target variable $Y$ and correlated among themselves. Precisely, $\rho$
is the correlation among any pair of the first $k$ variables, whereas
$\sqrt{\rho}$ and $\sqrt{\rho/2}$ are the level of correlation between
any relevant variable and, respectively, $f$ and $Y$. If we had an
infinite amount of observations, the sample correlation would be equal
to the true correlation between $X_j$ and $Y$. However, even in this
ideal asymptotic regime, this correlation would still remain a biased
indicator of the true relevance of each variable (represented by the
correlation between $X_j$ and $f$) due to the intrinsic noisy nature
of $Y$.

\begin{figure}[tp]
  \centering
  \vspace{-5mm}
  \includegraphics[width=0.35\textwidth]{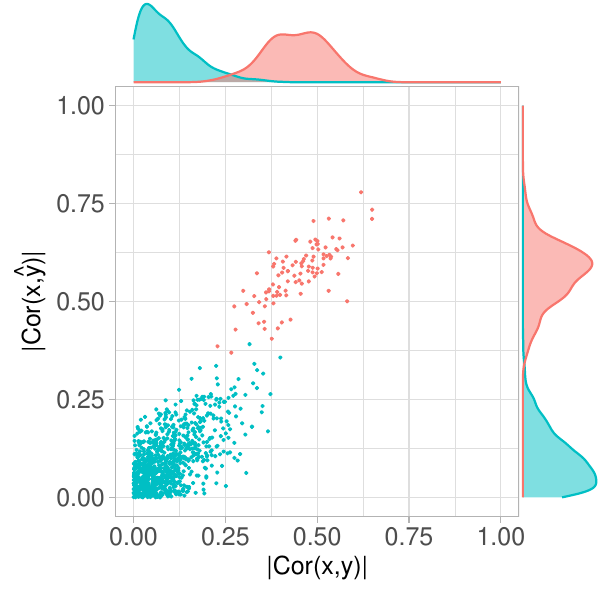}
  \vspace{-0.5\baselineskip}
  \caption{Sample correlation plot of each variable (relevant in red, 
  non-relevant in blue) with the target variable $y$ and the latent variable 
  $f$ respectively on the x and y axis. Simulated data are generated according to 
  \eqref{eq:simulated_data} with parameters $n=70$, $\rho=0.3$,
  $k=100$, $p=1000$.}
  \label{fig:correlation}
\end{figure}

When using a reference model, we first obtain predictions for $f$
using all the variables $\{X_{j}\}_{j=1}^{p}$ taking into account
that we have only observed the noisy representation $Y$ of $f$. If our
model is good, we are able to describe $f$ better than $Y$ itself can,
which improves the accuracy in the estimation of the relevance of the
variables.  Figure \ref{fig:correlation} illustrates this process in
the form of a scatter plot of (absolute) correlations of the
variables with $Y$ against the corresponding correlations with the
predictions of a reference model (in this case the posterior
predictive means of model \eqref{eq:ref_mod}; see Section
\ref{simulations}). Looking at the marginal distributions, we see
that using a reference model to filter out noise in the data, the two
groups of variables (relevant and non-relevant) can be distinguished
much better than when the correlation is computed using the observed
noisy data directly.

\hypertarget{comparison-minimal-subset}{
\section{Minimal subset variable selection}\label{comparison-minimal-subset}}

In the body fat example above, our two simultaneous goals were to 
obtain good predictive performance and to select a smaller number of variables. 
When the goal is to select a minimal subset of variables,
which have similar predictive performance as the full model, we call it
\emph{minimal subset variable selection}. This minimal subset might
exclude variables which have some predictive information about the
target but, given the minimal subset, these variables are not able to
provide such additional information that would improve predictive
performance in a substantial manner.
The usual reason for this is that the relevant variables
which are not in the minimal subset are highly correlated with variables already in
the minimal subset. We will return to a problem of finding all
the variables with some predictive power in Section~\ref{complete-selection}.

\subsection{Simulation study 1}

Using the data generating mechanism \eqref{eq:simulated_data}, we
simulate data sets of different sizes with a relatively large number
of variables $p=70$, with $k=20$ of them being predictive.  We compare
the minimal subset variable selection performance of the projection
predictive approach (which uses a reference model and it is referred to as projpred), 
a Bayesian stepwise
forward selection with and without a reference model, and maximum
likelihood stepwise forward selection with and without a reference
model (steplm). The following is a summary of the implementation of the compared methods:
\begin{itemize}
\item \textbf{projpred}: the projective prediction approach is used. 
  The reference model is a Bayesian linear regression
  model using the first five supervised principal components \citep{piironen2018} as predictors
  and the full posterior predictive distribution as the basis of the projection.   
  The search heuristic  is forward search and the predictive performance 
  is estimated via 10-fold cross-validation. The selection continues
  until the predictive performance is close to the predictive
  performance of the reference model. See more details in Appendix A.
  
\item \textbf{Bayesian stepwise selection (without a reference model)}: at each step, the fitted model
 is a Bayesian linear regression using the regularised horseshoe
  prior and the variable excluded is the one with the highest Bayesian
  $p$-value defined as $\text{min}\{P(\theta\leq0 \; | \; D),P(\theta>0 \; | \; D)\}$,
  where $D$ stays for the observed data.
The selection continues if the reduced model has an elpd score higher 
(i.e., better) than the current model.

\item \textbf{Bayesian stepwise selection (with a reference model)}: the reference model is the same as for projpred, but only point predictions (posterior predictive means) $\hat{y}$ are used to replace the target variable $y$. The same Bayesian
  $p$-value selection strategy as in the data based Bayesian stepwise selection is used.

\item \textbf{steplm (without a reference model)}: stepwise selection using AIC as in the body fat example.

\item \textbf{steplm (with a reference model)}: the reference model is the same as for
 projpred, but only point predictions (posterior predictive means) $\hat{y}$ 
 are used to replace the target variable $y$. The same AIC selection strategy 
 as in the data based steplm is used.

\end{itemize}

\begin{figure}[tp]
  \centering
  \includegraphics[width=0.98\textwidth]{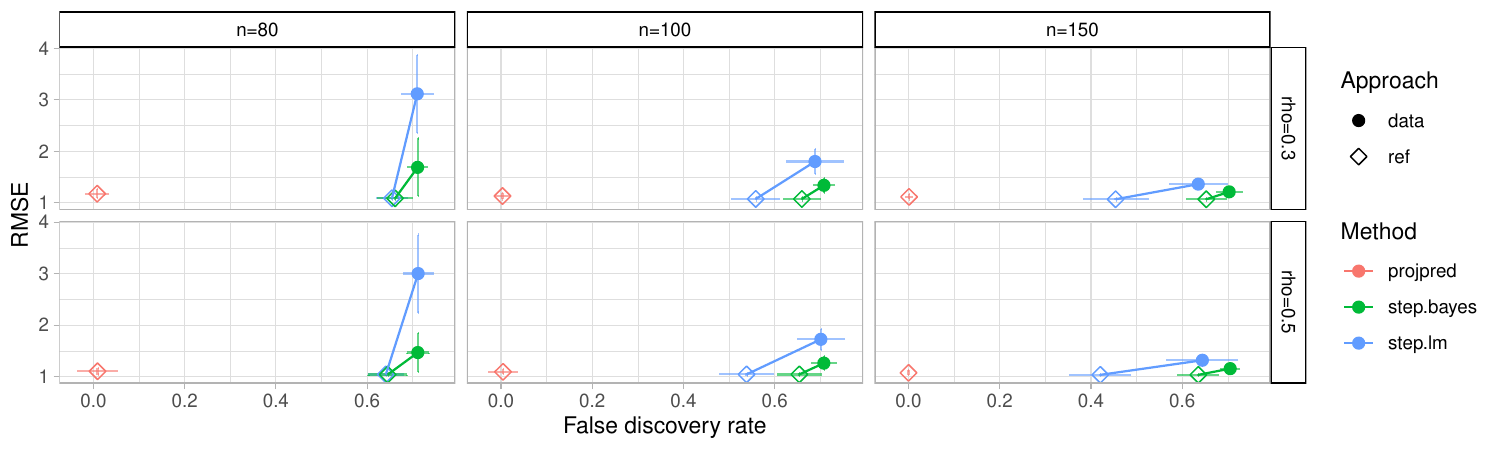}
  \vspace{-0.3\baselineskip}
  \caption{Simulation study 1: Root mean square error (RMSE) against false discovery rate in the minimal subset variable selection with one standard deviation error bars. The projpred approach has the smallest RMSE and false discovery ratio.}
  \label{fig:rmse_vs_fdr}
\end{figure}

In Figure \ref{fig:rmse_vs_fdr}, the predictive performance is shown
for different values of $n$ and $\rho$ in terms of RMSE and the false discovery rate (FDR, the ratio of the number of non-relevant selected
variables over the number of selected variables) of the selected submodel,
averaged after 100 data simulations. We see that projpred has superior FDR and the
smallest RMSE. Using a reference model improves steplm significantly,
but the minimal subset variable selection stays unreliable. Stepwise
Bayesian linear regression is better than steplm when comparing both methods with
and without a reference model, respectively. However, the minimal subset variable
selection of the Bayesian linear regression is still less reliable than projpred.

\begin{figure}[tp]
  \centering
  \includegraphics[width=0.98\textwidth]{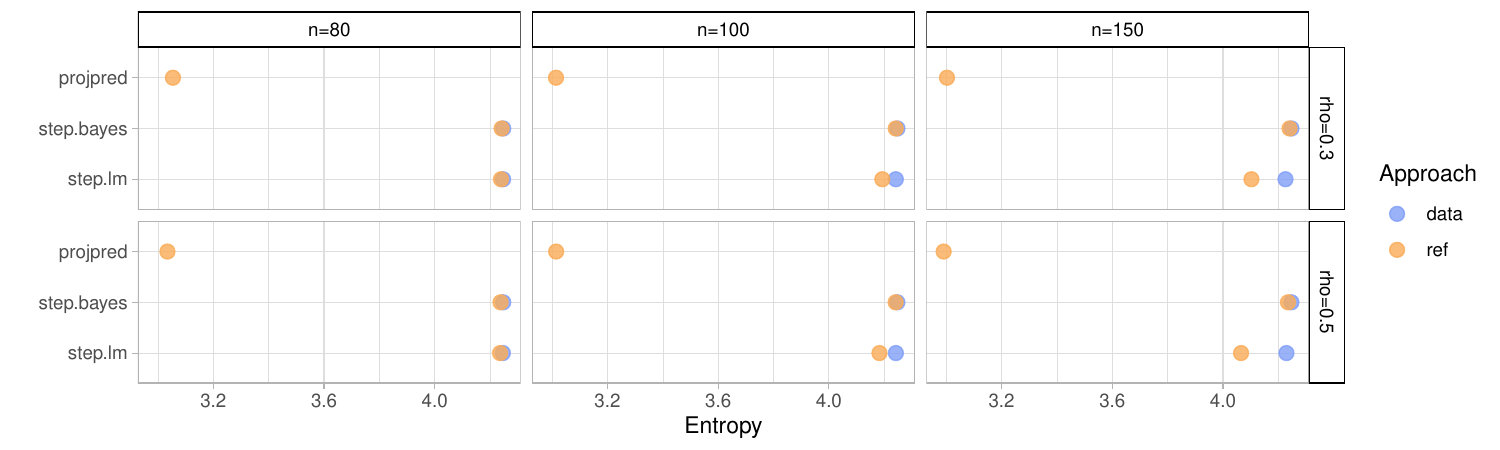}
  \vspace{-0.3\baselineskip}
  \caption{Simulation study 1: Entropy score in the minimal subset variable selection.  The projpred approach has much smaller entropy score than the other approaches.}
  \label{fig:entropy}
\end{figure}

When the variable selection is repeated with different simulated data
sets, there is some variability in the selected variables. We
measure the stability of variable selection by computing the entropy
of the observed distribution of the included variables over different models. The smallest entropy
would be obtained if the approach always selected the same set of
variables, and the largest entropy would be observed if the approach
would always selected different sets of variables. Therefore, lower entropy 
corresponds to a more stable selection. Highly correlated predictive variables may happen to be
selected alternately, thus making stability estimation of the
selection a non-trivial task. Entropy can not distinguish the interchangeability 
due to correlation from instability. Thus, such a measure should be 
considered as a relative, and not as an absolute, measure of stability.
Figure \ref{fig:entropy} shows the entropy scores for the different compared
methods. The use of a reference model
improves the stability of steplm in variable selection slightly, while it makes
little difference for the Bayesian linear regression. The
projpred approach turns out to be far more stable than all
other methods. This is likely due to projpred being based on better
decision theoretical formulation which 1) takes into account the full
predictive distribution and not just point estimate and 2) projects
the reference model posterior to the submodel instead of using a
simple refit of submodels.

\hypertarget{bodyfat-2}{
\subsection{Body fat example: Part 2}\label{bodyfat-2}}

\begin{figure}[tp]
  \centering
  \includegraphics[width=0.98\textwidth]{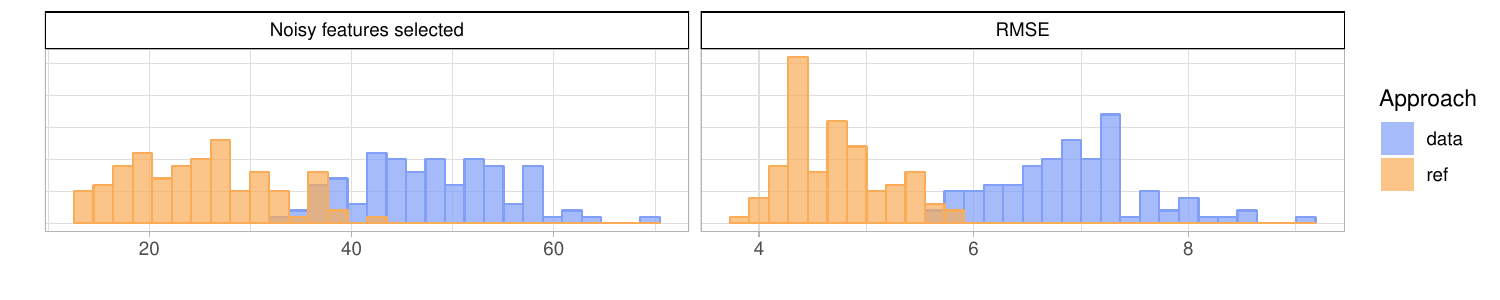}
  \vspace{-0.7\baselineskip}
  \caption{Body fat example: Stepwise backward selection with and without using a reference model. The x-axis denotes the number of selected irrelevant variables on the left and the out-of-sample RMSE on the right-hand side based on 100 bootstrap samples. The reference approach reduces the number of noisy variables selected and the out-of-sample RMSE.}
  \label{fig:bodyfat_step_refvsdata}
\end{figure}

Here we repeat the selection of Section \ref{bodyfat-1} via
stepwise backward regression. In this case, the overall number of
variables (original plus noisy) is 100, as it was in the last part of
Section \ref{bodyfat-1}. We compare results with and without using
a simple reference model approach outlined in \eqref{eq:ref_mod} with
steplm. Figure \ref{fig:bodyfat_step_refvsdata} shows the number of
irrelevant variables included in the final model and the out-of-sample
root mean square error (RMSE). Results are based on 100 bootstrap
samples on the whole dataset, and the predictive performance is tested
on the observations excluded at each bootstrap sample. We observe that
the reference model reduces the number of irrelevant variables included
in the final model. This leads to less overfitting and thus to
improved out-of-sample predictive performance in terms of RMSE. The
reference model approach applied to the stepwise backward regression
achieves outstanding improvements considering its simplicity, yet it
does not reach the goodness of the much more sophisticated projective
prediction approach (see results of Section \ref{bodyfat-1}).
\begin{table}[tp]
\footnotesize
\centering
\begin{tabular}{l|rr}
Method & RMSE & Selected noisy \\ 
  \hline
Stepwise selection & 6.9 (0.7) & 48 (7)    \\
Reference model + Stepwise selection & 4.7 (0.5) & 25 (7)   \\
\end{tabular}
\caption{Body fat example: Means and standard deviations (between brackets) of the results shown in Figure \ref{fig:bodyfat_step_refvsdata}.}
\label{tab:step_refvsdata}
\end{table}

\section{Complete variable selection}\label{comparison-complete-subset}
\label{complete-selection}

An alternative to minimal subset variable selection is \emph{complete
  variable selection} in which the goal is to find all relevant
variables that have some predictive information about the target. In
complete variable selection, it is possible that there are theoretically
relevant variables, but given finite noisy data we are not able to infer their relevance.
The projection predictive approach was originally designed for the minimal subset
variable selection but we will test a simple iterative variant for the complete
variable selection case in this Section.
 In addition, we analyse the benefits of the reference
model approach in combination with three other
methods which have been specifically designed for complete variable selection. 
As the criteria of selection performance, we evaluate the average false
discovery rate and the average
sensitivity (i.e., the ratio of the number of relevant selected
variables over the total number of relevant variables). We also provide
a comparison of the stability of the selection by means of a stability
measure proposed by \citet{paper:stability}, which goes from 0 to 1 and a higher value 
means a more stable selection. 

\subsection{Iterative projections}

We modify the projection predictive approach for complete variable
selection by using it iteratively.  Applying the straightforward
implementation of projpred, we are able to select a minimal subset of
variables, which yield to a model with a predictive performance
comparable to the full model's predictive performance. 
 The iterative projection repeats
the projpred selection for different iterations, at each time
excluding the variables selected in the previous
iterations from the search.
 At each iteration the selected submodel size corresponds to the
one having a predictive performance close enough to the baseline
model, which in this iterative version is the submodel with the
highest predictive score explored at the current iteration. This translates
in the following stopping rule at each iteration:
\begin{equation} 
\label{eq:rule_of_thumb}
\text{min} \{i\in \{0,\ldots,p\}: \text{P}(\text{elpd}_{i}-\text{elpd}_{\text{best}}>0)\geq \alpha \},
\end{equation}
where $i$ indexes the submodel size and ``best'' stands for the best predictive explored submodel
at the considered iteration. The algorithm terminates when the empty model (only intercept) satisfies
the stopping rule (see Algorithm \ref{alg:iterated_proj}).
The choice of the hyperparameter $\alpha$ is non-trivial, and we have 
observed sensitivity of the selection to such a choice, mainly when using cross-validation with a small
number of observations or not very predictive variables. 
In our experiments we chose the default value used in the \texttt{projpred} 
\texttt{R}-package, that is, $\alpha=0.16$.

\begin{center}
\begin{algorithm}[H]
\SetAlgoLined
\KwResult{$R:=\{\text{selected variables}\}$}
 $F:=\{\text{set of variables}\}$ \;
 $R=\{\emptyset\}$ \;
 Fit reference model\;
 \While{$F\neq\{\emptyset\}$}{
   projection.HeuristicSearch() \;
   projection.elpdEstimate() \;
   $S=\text{min}\{\text{sub}: P(\text{elpd}_{\text{sub}}-\text{elpd}_{\text{base}}>0)\geq\alpha)$\} \;
  
  \eIf{$S=\{\emptyset\}$}{
   break \;
   }{
   $R=R\cup S$\;
   $F=F\setminus S$\;
  }
 }
 \caption{Automated iterative projections}
 \label{alg:iterated_proj}
\end{algorithm}
\end{center}

In the experiments shown in the next sections, we include an additional iterative
method which we refer to as `iterative lasso'. It consists of the same iterative
algorithm as iterative projpred expect for not using any reference model, but
the lasso method for variable selection, instead.
That is, it uses the observed target values instead of predictions of the reference model.
The comparison with iterative lasso can help to disentangle the effects
of the iterative procedure and the usage of a reference model in complete feature selection.

\subsection{Alternative complete variable selection methods}

We consider three alternative complete selection methods:
 the control of the local
false discovery rate \citep{paper:efron, efron2012large}, the
empirical Bayes median \citep{johnstone2004needles}, and the selection
by posterior credible intervals.

The control of the local false discovery rate consists of testing the
z-values $\{z_{j}\}_{j=1}^{p}$ of a normal mean problem (explained in Section \ref{simulations}) 
on whether they belong to the
theoretical null distribution $f_{0}$ (i.e., the null hypothesis $H_0$
meaning no relevance) against the alternative hypothesis distribution
$f_{1}$. In our case $f_{0}$ corresponds to the standard normal
distribution (see expression \eqref{eq:normal_means_problem2}). The
quantity of interest is the local false discovery rate (loc.fdr)
defined as: \
\begin{equation}
\text{loc.fdr}(z)=P(H_{0}|z)=\frac{f_{0}(z)\pi_{0}}{f(z)},
\end{equation}
where $\pi_{0}$ is the prior probability of $H_0$ and
$f(z)=\pi_{0}f_{0}(z)+(1-\pi_{0})f_{1}(z)$ is the marginal
distribution of the z-values. The latter is estimated using splines
with 7 degrees of freedom. We select variables with local false
discovery rate below $0.2$, which is suggested by
\citet{efron2012large} as it corresponds to a Bayes factor larger than
36 (assuming $\pi_{0}\geq0.9$). The results of the comparison are not
sensitive to the specific value. To estimate $\pi_{0}$ from the data,
we use the default setting provided by the \texttt{R}-package
\texttt{locfdr} \citep{Efron+Turnbull+Narasimhan:2015:locfdr}.

The empirical Bayes median approach consists of fitting a Bayesian
model with a prior composed by a mixture of a delta spike in zero and
a heavy-tailed distribution. We use the implementation in the
\texttt{R}-package \texttt{EbayesThresh}
\citep{Silverman+etal:2017:EbayesThresh}. As suggested by
\cite{johnstone2004needles}, we use a Laplace distribution resulting
in a thresholding property, that is, there exists a threshold value
such that all the data under that threshold have posterior median
equal to zero. Therefore the selection is done by selecting only those
parameters whose posterior median is different from zero. The
hyperparameter of the Laplace distribution and the mixing weight of
the prior are estimated by marginal maximum likelihood.

The selection
by 90\% posterior credible intervals is done using the regularised
horseshoe prior \Citep{paper:rhs} and selecting those variables whose posterior
distribution does not include zero in the interval between the 5\% and
the 95\% quantiles.

All of these methods provide a complete selection approach and we
compare their performance with and without using a reference model.
That is, in the data condition, we apply the method on the original
data $y$ while, in the reference model condition, we replace $y$ by
their mean predictions $\hat{y}$ based on the reference model.

\subsection{Simulation study 2}
\label{simulations}

The iterative projection applies straightforwardly to data, whereas
to investigate the performance of the three alternative complete selection approaches,
we are going to use simulations based on the normal means problem.
The normal means problem consists of estimating the (usually
sparse) vector of means of a vector of normally distributed
observations. The dimensionality of the vector of means is denoted by
$p$ and $\{z_{j}\}_{j=1}^{p}$ is the vector of observations of the
random variables $\{Z_{j}\}_{j=1}^{p}$. The task is to estimate the
latent variables $\{\theta_{j}\}_{j=1}^{p}$ of the following model: \
\begin{equation}\label{eq:normal_means_problem}
Z_{j}|\theta_{j},\sigma\overset{ind}{\sim}\N(\theta_{j},\sigma^{2}), \quad j=1,\ldots,p.
\end{equation}
This is equivalent to a linear regression where the design matrix is
the identity matrix with the number of variables being equal to the
number of observations. This formulation can be found in practice, for
example, in the analysis of microarray data, where a large set of
genes are tested in two groups of patients labeled as positive or
negative to some disease \citep{paper:efron,efron2012large}. The
objective of the analysis is to select the subset of genes
statistically relevant to the disease. One common way to proceed is to
compute the two-sample $t$-statistic for every gene separately.  After
normalising these statistics, they then become the data $Z$ in the
normal means problem \eqref{eq:normal_means_problem}. For further
details see the examples by \cite{paper:efron, efron2012large}.

In our experiments, we retrieve the normal means problem from the
sample correlations between the target and the variables using the
Fisher $z$-transformation \citep{hawkins1989using}. Suppose we have a
continuous target random variable $Y,$ a set of $p$ continuous
variables $\{X_{j}\}_{j=1}^{p}$, and denote
$\rho_{j}=\Cor(Y,X_{j})$. Further, suppose we have observed $n$
statistical units and define $r_{j}$ the sample correlation between
the observations of the target variables $\{y_{i}\}_{i=1}^{n}$ and the
$j$-th variable $\{x_{ij}\}_{i=1}^{n}$. Finally, we refer to the
Fisher $z$-transformation function $\text{tanh}^{-1}(\cdot)$ as
$T_{F}(\cdot)$. Assuming each pair $(Y,X_{j})$ to be bivariate
normally distributed, the corresponding transformed correlations are
approximately normally distributed with known variance: \
\begin{equation} \label{eq:fisher_transformation}
T_{F}(r_{j})\overset{ind}{\sim} \N\left(T_{F}(\rho_{j}),\frac{1}{n-3}\right), \quad j=1,\ldots,p.
\end{equation}
Therefore, rescaling the quantities $T_{F}(r_{j})$ by $\sqrt{n-3}$ and
denoting the results as $z_{j}$, we have the formulation
\eqref{eq:normal_means_problem} of the normal means problem, this time
with unit variance: \
\begin{equation} \label{eq:normal_means_problem2}
Z_{j}|\theta_{j}\overset{ind}{\sim}\N(\theta_{j},1), \quad j=1,\ldots,p.
\end{equation}
In this case, the quantities of interest $\theta_{j}$ are equal to
$\sqrt{n-3}\,T_{F}(\rho_{j})$.

In our simulations, we use different levels of correlation 
$\rho\in\{0.3,0.5\}$ and numbers of
observations $n\in\{50,70,100\}$. The total number of variables $p$
and the number of relevant variables $k$ are fixed to $p = 1000$ and
$k = 100$, respectively. In general, the lower $\rho$ and $n$, the
more challenging the variable selection is. For this example, \citet{paper:projpred} proposed to
use a reference model which a Bayesian linear regression using the first five
supervised principal components (SPC) as variables and imposing an hierarchical prior on
their coefficients:
\begin{equation}
\label{eq:ref_mod}
\begin{aligned}
    Y_{i}|&\boldsymbol{\beta},\sigma^{2},\boldsymbol{u}_{i} \overset{ind}{\sim} \N(\boldsymbol{u}_{i}^{T}\boldsymbol{\beta},\sigma^{2}) \quad &i=1,\ldots,n \\
    &\beta_{j}|\!\begin{aligned}[t] &\tau \overset{iid}{\sim} \N(0,\tau^{2})\\
    &\tau \sim t_{4}^{+}(0,s_{max}^{-2}) 
    \end{aligned} &j=1,\ldots,5 \\ 
    &\sigma \sim t_{3}^{+}(0,10). \\
\end{aligned}
\end{equation}
In the above, $u_{ij}$ represents the $j$-th SPC evaluated at
observation $i$, and $s_{max}$ denotes the sample standard deviation
of the largest SPC.  The SPCs are computed using the
\texttt{R}-package \texttt{dimreduce} (\url{https://github.com/jpiironen/dimreduce}) setting the screening threshold
parameter at $0.6s_{max}$.  In our experiments, the results are not
sensitive to the specific choice of the screening threshold, yet a more
principled approach would be to use cross-validation to select the
threshold as done by \citet{paper:projpred}.

\begin{figure}[tp]
  \centering
  \includegraphics[width=0.98\textwidth]{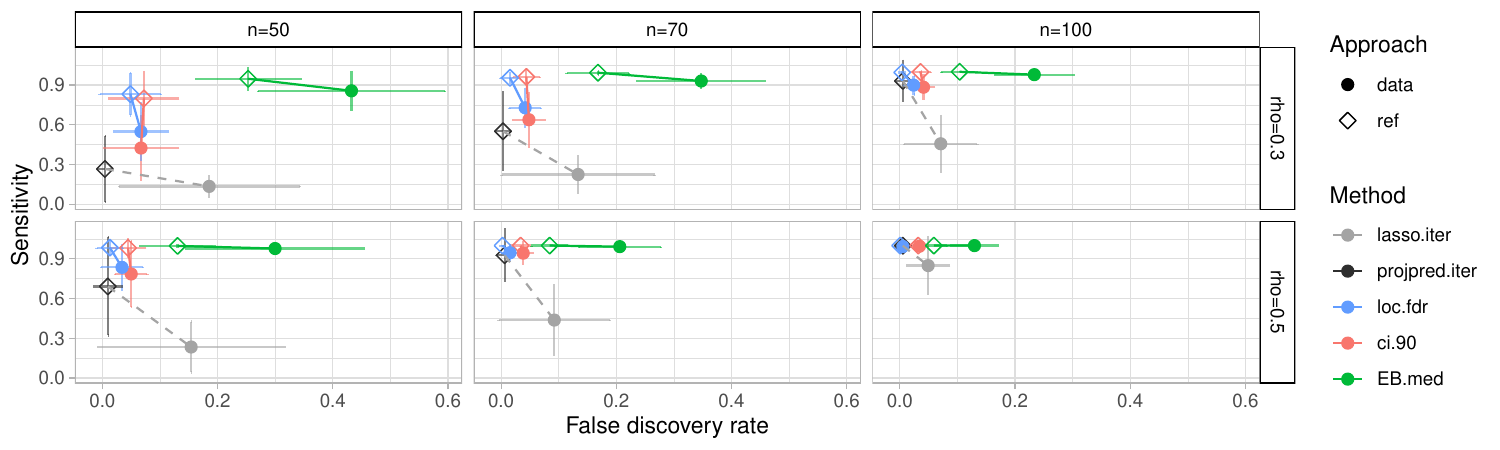}
  \vspace{-0.3\baselineskip}
  \caption{Simulation study 2: Complete variable selection sensitivity against false discovery rate based on 100 data simulations with one standard deviation error bars. The reference approach improves sensitivity and reduces false discovery rate for all methods.}
  \label{fig:sensitivity_vs_fdr_iterated}
\end{figure}
\begin{figure}[tp]
  \centering
  \includegraphics[width=0.98\textwidth]{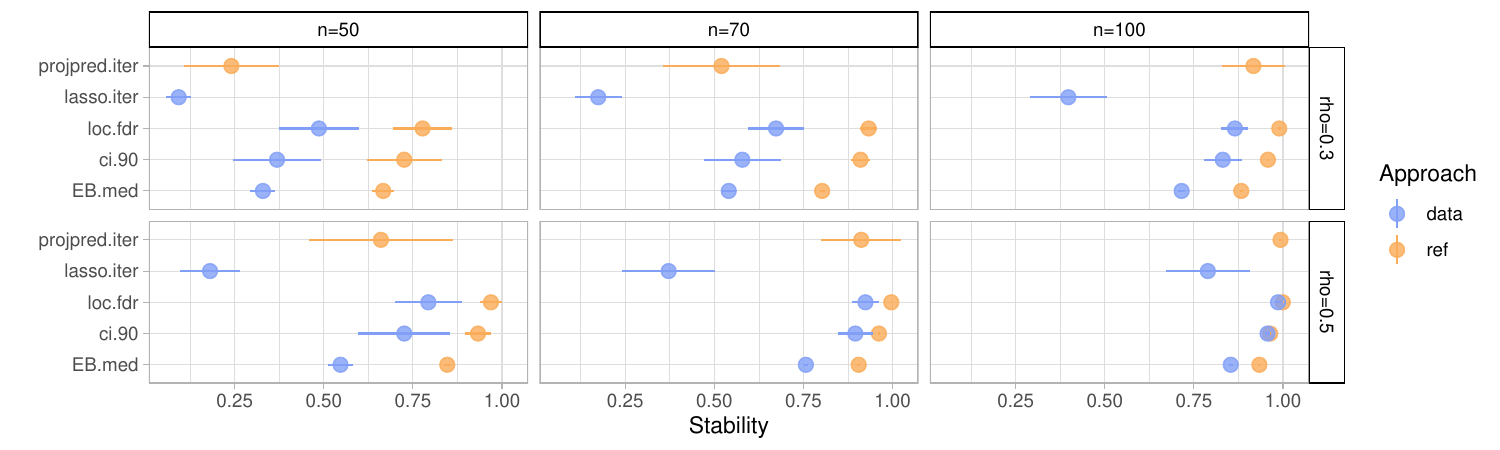}
  \vspace{-0.3\baselineskip}
  \caption{Simulation study 2: Complete variable selection stability estimates with 95\% intervals based on 100 data simulations. The reference approach improves stability for all methods.}
  \label{fig:stability_iterated}
\end{figure}

Figure \ref{fig:sensitivity_vs_fdr_iterated} shows the average sensitivity on
the vertical axis and the average false discovery rate on the
horizontal axis based on 100 data simulations for the different
combinations of $n$ and $\rho$. The best selection performance is on the
top-left corner of each plot, as it implies the lowest false discovery
rate and the highest sensitivity. We see that regardless of the
applied selection method, the use of a reference model improves the
selection performance, as it reduces the false discovery rate (shifting
to the left) or increases the sensitivity (shifting upwards). In
accordance with what can be expected, the larger data set size ($n$) and the higher the true correlations ($\rho$), the easier the
selection is. Thus, for easier selection scenarios, the benefits of
the reference model are smaller since the raw data already provide
enough information to identify the relevant variables. The iterative
projpred has good false discovery rate in all cases, and the
sensitivity is good except when the number of observations and the
correlation level are small. It performs better than the iterative lasso 
selection in any of the simulated scenarios. Tuning $\alpha$ might lead to improved
selection, yet the main source of poor performance could well be the
miscalibration of the uncertainty of the predictive utility score for
misspecified models with cross-validation \citep{bengio2004no}.

Figure \ref{fig:stability_iterated} shows the estimates of the stability
measure proposed by \cite{paper:stability} with 0.95 confidence
intervals based on 100 simulations. Such a measure takes in account
the variability of the subset of the selected variables at each
simulation (originally at each bootstrap sample), modelling the
selection of each variable as a Bernoulli process. Further details are
available in \cite{paper:stability}. The reference model helps in
improving the stability of the selection: again, the benefits are
bigger when the problem is more difficult (small $n$ and $\rho$). In
addition, we observe less uncertainty in the stability estimates for
the reference approach (i.e., smaller width of the 95\% intervals),
which can be still connected to the overall stability of the
procedure. As in Figure \ref{fig:sensitivity_vs_fdr_iterated}, the iterative
projection does not perform well in the hardest scenarios.

\hypertarget{bodyfat-2}{%
\subsection{Body fat example: Part 3}\label{bodyfat-3}}

\begin{figure}[tp]
  \centering
  \includegraphics[width=0.98\textwidth]{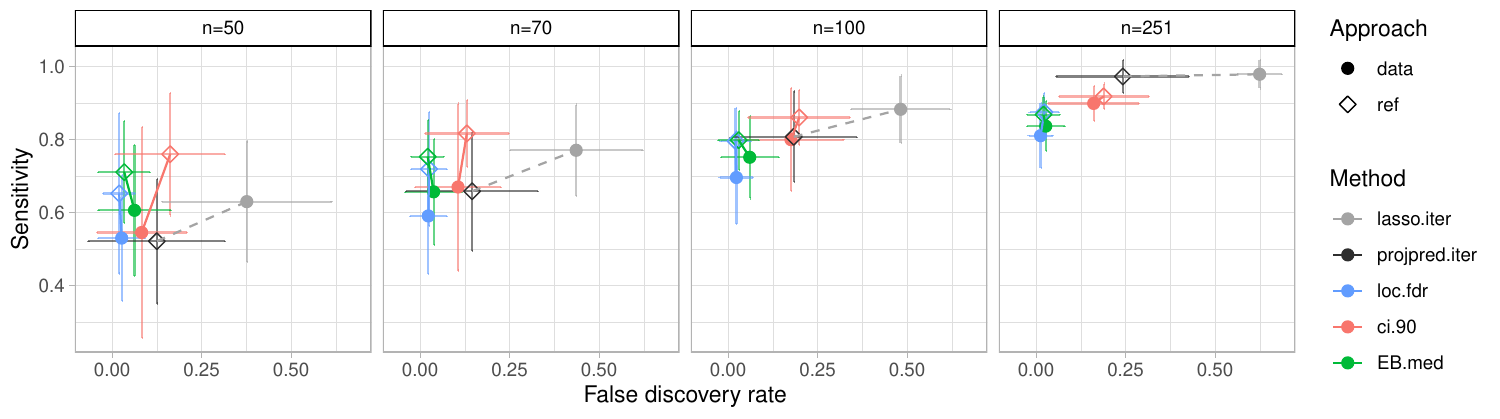}
  \vspace{-0.3\baselineskip}
  \caption{Body fat example with noisy variables: Complete variable selection sensitivity against false discovery rate based on 100 bootstrap samples with one standard deviation error bars. The improvement from using the reference approach is small (except that the projpred is much better than lasso).}
  \label{fig:bodyfat_sensitivity_vs_fdr}
\end{figure}
\begin{figure}[tp]
  \centering
  \includegraphics[width=0.98\textwidth]{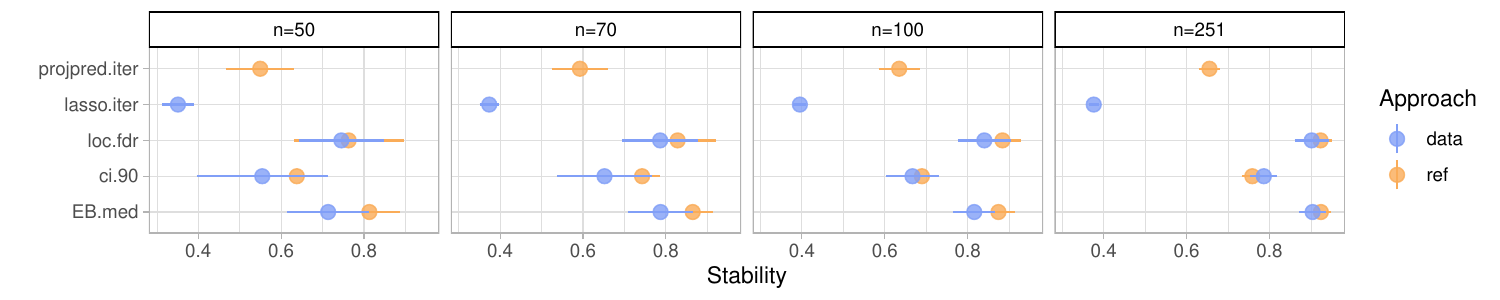}
  \vspace{-0.3\baselineskip}
  \caption{Body fat example with noisy variables: Complete variable selection stability estimates with 0.95 confidence intervals based on 100 bootstrap samples. The improvement from using the reference approach is small (except that the projpred is much better than lasso).}
  \label{fig:bodyfat_stability}
\end{figure}

We conclude our complete selection experiments using the body fat
dataset one more time. As earlier we add noisy uncorrelated variables
to the original data to get a total of 100 variables.
Since we do not have a ground truth available with regard the original
variables of the data, we assume it is reasonable to consider all of
them relevant, at least to some degree. The artificially added
variables are naturally irrelevant by construction.
We compute
correlations between each variable and the target variable, that is
the amount of fat, and transform them by Fisher-Z-transformation. The
original assumption in order for \eqref{eq:fisher_transformation} to
hold is that the variables are jointly normally distributed. In our
experience the normal approximation in
\eqref{eq:fisher_transformation} is still reasonable, but after
rescaling by $\sqrt{n-3}$ we do not fix the variance to be one, and
instead estimate it from the data. We compare the iterative projection, 
the control of the local false discovery rate (loc.fdr), 
the empirical Bayes median (EB.med)
and the selection by posterior credible intervals at level 90\%
(ci.90).  In order to vary the difficulty of the selection, we
bootstrap subsamples of different sizes, going from $n=50$ up to
$n=251$ (i.e., the full size of the data). For each condition, results
are averaged over 100 bootstrap samples of the respective size.

Figure \ref{fig:bodyfat_sensitivity_vs_fdr} shows the sensitivity
against the false discovery rate.  In almost all of the bootstrapped
subsamples, the reference
model improves the selection both in terms of sensitivity and false
discovery rate. When $n=50$, we observe worse false
discovery rates, yet by a lower amount compared to the gain in
sensitivity. Again, we observe that the benefits are more evident as
the selection becomes more challenging (i.e., lower number of
observations).
The great performance of projpred in minimal subset selection is not
carried over for the complete variable selection with iterative
projpred (even changing the reference model with a full encompassing 
linear regression with regularised horseshoe prior), and the methods 
specifically designed for the complete variable selection perform better. 
However we still observe a better selection with respect to iterative lasso
in any of the examined scenarios, mainly in terms of false discovery rate.
Figure \ref{fig:bodyfat_stability} shows the stability
results using the measure by
\cite{paper:stability}. The benefits of the reference model are here
marginal with only small improvements.

In this example, we have used the reference model defined as a linear
regression over some supervised principal components, because it is
natural for a large number of correlating variables, and has fairly good
predictive performance plus it is computationally efficient. We do not
argue that this is always the best choice and more sophisticated
models can lead to even better results. Here the purpose of the
experiments were to motivate the use of reference models in general,
and as we needed to average results over a lot of repetitions per
simulation condition, we preferred such a comparably simple and
computationally fast reference model.

\hypertarget{conclusion}{
\section{Conclusion}\label{conclusion}}

In this paper, we demonstrated the benefits of using a reference model
to improve variable selection, or more generally, model reduction. 
We have motivated and explained the general benefits
of a reference model regardless of the method it is applied in combination
with. Specifically, we have seen how the reference model acts as an
approximation of the data generation mechanism through its predictive
distribution. Such approximation is generally less noisy than the
sample estimation available purely from the observed data, leading to
the main benefits of the reference model approach. In our comparisons,
we have analysed the effect of a reference model in the form of a
filter on the observed target values on top of different widely used
variable selection methods. Overall, using a reference model leads to
more accurate and stable selection results independently of the
specific selection method. These benefits apply to a large family of
different methods all involving a reference model in one way or the
other. Some of these approaches have been present in the literature for some time
\cite[e.g., see references in][]{vehtari2012survey,paper:projpred}
but often without a clear explanation of {why} they are actually
favourable and how they connect to other related approaches. We hope
that the present paper can fill some of these gaps by providing a
unifying framework and understanding of reference models.

We argue that, whenever it is possible to construct a reasonable
reference model, it should be employed on top of the preferred
selection procedure or as an integral part of more complex methods,
for example, the projective prediction approach
\citep{paper:projpred}. Note that one of the main challenges in many
real world application will consist in devising a sensible reference
model itself and assessing its predictive performance.
To build good predictive reference models, which are specifically
tuned to the data and problem at hand, we recommend them to be
developed using a robust Bayesian modelling workflow, for instance, as
outlined by \citet{gelman2013bayesian} and \citet{gabry2019visualization}.

Another main result of this paper is that the projective prediction
approach shows superior performance in minimal subset variable
selection compared to alternative methods whether or not these methods
make use of a reference model. That is, while the reference model is
certainly one important aspect the projective prediction approach, it
is not the only reason for its superior performance.  Rather, by
incorporating the full uncertainty of the posterior predictive
distribution into the variable selection procedure (instead of just
using point estimates) and using principled cross-validation method,
projective predictions combine several desirable variables into a
single procedure \citep{paper:projpred}.  In summary, we would
strongly recommend using projective predictions for minimal subset
 variable selection if possible and feasible. 
 However, if this is not an option in a given
situation, we would in any case recommend using a reference model on
top of the chosen variable selection method.

The projective prediction approach was not designed for the complete
variable selection. We tested a simple iterative version of projpred,
with mixed results and the methods specifically designed for complete
variable selection (especially loc.fdr) performed better in our
experiments. It is left for future research to develop 
a better projective prediction approach for complete variable
selection problems.

All Bayesian models in this paper have been implemented in
the probabilistic programming language \texttt{Stan}
\citep{paper:stan} and fit via dynamic Hamiltonian Monte Carlo
\citep{hoffman2014no,betancourt2017conceptual}, through the \texttt{R}-packages 
\texttt{rstan} \citep{Rrstan} and \texttt{rstanarm} \citep{Rrstanarm}.
Graphics elaborations have been done using \texttt{ggplot2} \citep{Rggplot2}
and the \texttt{tidyverse} framework \citep{Rtidyverse}.
The code to run all
the experiments is available on GitHub
(\url{https://github.com/fpavone/ref-approach-paper}).

\section*{Acknowledgments}

We thank Alejandro Catalina Feliu for help with experiments, and
Academy of Finland (grants 298742, and 313122), Finnish Center for
Artificial Intelligence and Technology Industries of Finland
Centennial Foundation (grant 70007503; Artificial Intelligence for
Research and Development) for partial support of this research. We
also acknowledge the computational resources provided by the Aalto
Science-IT project.
 
\bibliography{ref_approach}

\appendix

\section*{Appendix}

\subsection*{Appendix A: Projective Predictions}

The projective prediction (projpred) approach was developed and is thoroughly
 described by \cite{paper:projpred}. In this appendix, we provide a high level
  description of the method so that readers do not need to study paper by \cite{paper:projpred} in detail
to understand the main ideas behind projpred.

The parameter
distribution of a given candidate submodel is denoted by $\pi$ and the induced predictive
distribution by $q_{\pi}(\tilde{y})$. We would like to choose $\pi$ so
that $q_{\pi}(\tilde{y})$ maximises some predictive performance
utility, for example, the expected log-predictive density (elpd)
defined as: \
\begin{equation}\label{eq:elpd}
\text{elpd}[q_{\pi}]=\int \text{log}\,q_{\pi}(\tilde{y})p_{t}(\tilde{y})d\tilde{y},
\end{equation}
where $p_{t}(\tilde{y})$ denotes the (usually unknown) true generating
mechanism of future data $\tilde{y}$. If we refer to the posterior
predictive distribution of a reference model with $p(\tilde{y}|D)$,
where $D$ stands for the data on which we conditioned on, we can
approximate \eqref{eq:elpd} using $p(\tilde{y}|D)$ instead of the true
data generation mechanism $p_{t}(\tilde{y})$. The maximisation of the
elpd using the reference model's predictive distribution is equivalent
to the minimisation of the Kullback-Leibler (KL) divergence from the
reference model's predictive distribution to the submodel's predictive
distribution: \
\begin{equation} \label{proj_as_filter}
\underset{\pi}{\text{max}} \; \int \text{log}\,q_{\pi}(\tilde{y})p(\tilde{y}|D)
\, \text{d} \tilde{y} 
\quad \Leftrightarrow \quad 
\underset{\pi}{\text{min}} \; \text{KL}[p(\tilde{y}|D) \; || \; q_{\pi}(\tilde{y})] 
\end{equation}
The term on the right-hand side of Equation \eqref{proj_as_filter}
describes what is referred to as the projection of the predictive
distribution, which is the general idea behind the projection
predictive approach \cite[see][]{paper:projpred}. 
We now summarise the workflow of the projection predictive approach in
the particular case of the draw-by-draw projection \cite[original
formulation by][]{paper:original_proj}, following
\cite{paper:projpred}. Suppose we have observed $n$ statistical units
with target values $\{y_{i}\}_{i=1}^{n}$ and a set of observed
variables for which we want to obtain a minimally relevant
subset. Than, the main steps are the following:
\begin{enumerate}
\item Devise and fit a reference model. Let $\{\boldsymbol{\theta}_{*}^{s}\}_{s=1}^{S}$ be the set of $S$ draws from the reference model's posterior.
\item Rank the variables according to their relevance using some
  heuristics and consider as candidate submodels only those which
  preserve this order, starting from including only the highest ranked
  variable. The submodels are then naturally identified by their
  model size. This step is not strictly necessary but reduces the number
  of submodels to considered in the following steps and thus reduces computation time.
\item For each submodel $\pi$ selected in Step 2, project each of the reference model's posterior draws $\boldsymbol{\theta}_{*}^{s}$ as follows:
\begin{equation}\label{eq:draw_by_draw}
\boldsymbol{\theta}_{\perp}^{s} = \underset{\boldsymbol{\theta^{s}}\in\Theta}{\text{argmin}} \frac{1}{n}\sum_{i=1}^{n}\text{KL} \left[ p(\tilde{y}_{i}|\boldsymbol{\theta}_{*}^{s})\;||\;q_\pi(\tilde{y}_{i}|\boldsymbol{\theta}^{s}) \right],
\end{equation}
where $p(\tilde{y}_{i}|\boldsymbol{\theta}_{*}^{s})$ stands for the
predictive distribution of the reference model with parameters fixed
at $\boldsymbol{\theta}_{*}^{s}$ and conditioning on all the variable
values related to the statistical unit (identified by the subscript
$i$), whereas $q_\pi(\tilde{y}_{i}|\boldsymbol{\theta}^{s})$ is the
predictive distribution of the submodel. The projected draws
$\boldsymbol{\theta}_{\perp}^{s}$ then present the projected posterior
for the submodel.
\item For each submodel (size), test the predictive performance for a chosen predictive utility score, for example, via cross-validation. Fast cross-validation can be performed using using approximate leave-one-out cross-validation via Pareto-smoothed importance-sampling
\citep[PSIS-LOO-CV; ][]{paper:psis_loo}.
\item Choose the smallest submodel (size) that is sufficiently close to the reference model's predictive utility score. The results in this paper were not sensitive to the specific choice of how ``sufficiently close'' is defined, and we used the same definition as \citet{paper:projpred}.
\end{enumerate} 

In general, Expression \eqref{eq:draw_by_draw} is not an easy optimisation problem
However, in the special case of the submodels being
generalised linear models with a likelihood coming from the exponential family, \eqref{eq:draw_by_draw} reduces to a
maximum likelihood estimation problem, which can be easily solved
\citep{paper:original_proj}. For further details on the projective
prediction workflow and implementation see the paper by \citet{paper:projpred}.

\end{document}